# Loop Free Multipath Routing Algorithm

Rashmi Singh
Electrical Engineering Department
Indian Institute of Technology, Kanpur
Kanpur, India 208016
Email: rashmi0888@gmail.com

Yatindra Nath Singh
Electrical Engineering Department
Indian Institute of Technology, Kanpur
Kanpur, India 208016
Email: ynsingh@iitk.ac.in

Anita Yadav
Computer Science and Engineering Department
Harcourt Butler Technological Institute
Kanpur, India 208002
Email: anitacse7@gmail.com

*Abstract*—Single path routing that is currently used in the internet routers is easy to implement as it simplifies the routing tables and packet flow paths. However it is not optimal and has shortcomings in utilizing the network resources optimally, load balancing & fast recovery in case of faults (fault tolerance). The given algorithm resolves all these problems by using all possible multiple paths for transfer of information, while retaining loop-free property. We have proposed a new dynamic loop-free multipath routing algorithm which improves network throughput and network resource utilization, reduces average transmission delay, and is not affected by faults in the links and router nodes. The main idea of this algorithm is to maintain multiple possible next hops for a destination along with weights. At every node, the traffic to a destination is split among multiple next hops in proportion to the estimated weights. The number of multiple next hops also changes depending on the traffic conditions, but it is never less than one.

## I. Introduction

Since its birth decades ago, Internet has changed life of people by allowing data to be transacted and thus has brought about the most significant revolution in communication. Many applications like e-mail, web browsing have become ubiquitous today. Further real-time services like IP telephony, IPTV, Video conferencing are taking their share in internet traffic. The growth in both the number of users of the internet and in their bandwidth and quality requirements has placed increasing demands on the ISPs networks. To meet these demands and to respond to the challenges, performance optimization of networks is required which is accomplished by routing traffic in a way to utilize network resources efficiently and reliably.

The routing algorithms used in today's computer networks and internetworks typically focus on discovering a single optimal path for routing, according to some desired routing metric which can be based on distance, delay, bandwidth, reliability or a combination of them. Accordingly, traffic is always routed over a single path. While the simplicity of this approach has made IP routing highly scalable, it often results in substantial waste of network resources.

Multipath routing is fundamentally more efficient than the currently used single-path routing protocols. It is an effective strategy to achieve robustness, load balancing, reduction in congestion and end to end delay, and it achieves all of these by distributing load across multiple paths. The provision of multiple paths in this type of routing makes it fault-tolerant, reliable and increases network throughput. It also prevents network oscillations by avoiding congestion of links.

There are two aspects of any multipath routing algorithm: computation of multiple loop-free paths and traffic splitting among these multiple paths. None of these aspects are exploited to their full potential in current protocols. For example, OSPF-ECMP [1] allows a router to choose more than one path to the same destination only when those paths offer the minimum distance. But when there is fine granularity in link costs metric, as in the case of optimal routing, there is less likelihood that multiple paths with equal distance exist between each source-destination pair. OSPF-OMP [2] somewhat relaxes this best path criteria to allow a neighbor node closer in terms of cost to the destination than the current node to be a viable next hop. But still it does not utilize the full connectivity of underlying physical network in which any source-destination pair might be connected by unequal cost multipaths.

Splitting of traffic among the multiple paths in OSPF is also not very optimal which further limits the ability to decrease congestion through load balancing. For e.g., in OSPF-ECMP [1] load is distributed equally over multiple equal-cost paths. But to make optimal use of network resources and minimize delays, traffic between source-destination pairs may often have to be split and routed along multiple paths in proportions that are not necessarily equal. Though OSPF-OMP [2] suggests using unequal traffic distribution on multiple paths, the distribution is based on a heuristic scheme that often results in inefficient flow distribution.

Several existing and proposed routing protocols are designed to use unequal-cost paths to a destination. An approximation to minimum -delay routing is presented in [3]. DASM [4] and MDVA [5] determine the set of feasible next-hops based on the minimum distance through each neighbor. MPATH [6] is a path finding algorithm that uses distance vectors combined with the identity of the second-to-last node, also called predecessor node, that is just before the destination on the shortest path to find the feasible next hops. EIGRP [8] selects as next-hops the neighbors with a distance to the destination that is no longer than the shortest path times a variance factor. The above mentioned algorithms suffer from some drawbacks. Since they build multiple paths to a destination all the time, they increase complexity of system as the routing tables grow in size and also generate additional overhead in maintaining and reconfiguring multiple routes. Also, under low traffic conditions, splitting of traffic among multiple paths does not offer any significant performance improvement over single paths. Also, none of the above mentioned routing algorithms clearly mentions the traffic split policy among the chosen multiple paths.

We have presented a new loop-free distance-vector algorithm which maintains multiple possible next hops for a destination along with weights and hence provides load balancing even for unequal cost multiple paths, splits the traffic to a destination among multiple next hops in proportion to the estimated weights and thus provides a more efficient flow distribution, changes the number of next hops at each node based upon the traffic conditions and thus has a lower complexity than pure multipath routing algorithms.

The paper is organized as follows. Section II discusses the implementation basics of our proposed loop-free multipath routing algorithm. Section III gives the algorithm. Section IV presents results of simulation experiments designed to illustrate the comparative benefits of our algorithm over other single-path and multipath routing algorithms. Section V concludes the paper.

## II. Loop Free Multipath Routing Algorithm

The algorithm proposes a new methodology for selecting the number of paths between any source-destination pair and calculating traffic split ratios for them. The key features of the algorithm that enables it to achieve this is the computation of adaptive cost metric to a destination and the formation of forward and backward sets at every node for every destination. We describe, in the following subsections, these key features of the algorithm along with the routing tables structure, various packet types and timers it supports.

### A. Forming Forward and Backward Sets

Consider a node $k$. For a destination $d$, set $N_k$ of all the neighboring nodes of $k$, is partitioned into two sets – forward and backward nodes of $k$. The set of neighbors in the forward direction is $F_d^k$ and in the backward direction is $B_d^k$ (Fig. 1), then following holds:

- The set $F_d^k$ is nothing but multiple possible next hops from node $k$ to destination $d$.
- If a node $i$ is in set $F_d^k$ of node $k$, then $k$ will be in set $B_d^i$ of $i$. If $i$ is moved to set $B_d^k$ of $k$, then $k$ will also be moved to set

TABLE I: NOTATION

| $N$ | Set of nodes in the network |
|---|---|
| $N_k$ | Set of neighbors of node $k$ |
| $d$ | Destination |
| $F_d^k$ | Set of neighbors in forward direction for $d$ at node $k$ |
| $B_d^k$ | Set of neighbors in backward direction for $d$ at node $k$ |
| $C_{ki}$ | Capacity of link $(k,i)$ |
| $C_{ki}^d$ | Capacity from $k$ to $d$ via $i$ |
| $C_k^d$ | Overall capacity from $k$ to $d$ |
| $u_{ki}$ | Utilization of link $(k,i)$ |
| $r_i^d$ | Load split ratio on link to node $i$ for destination $d$ |

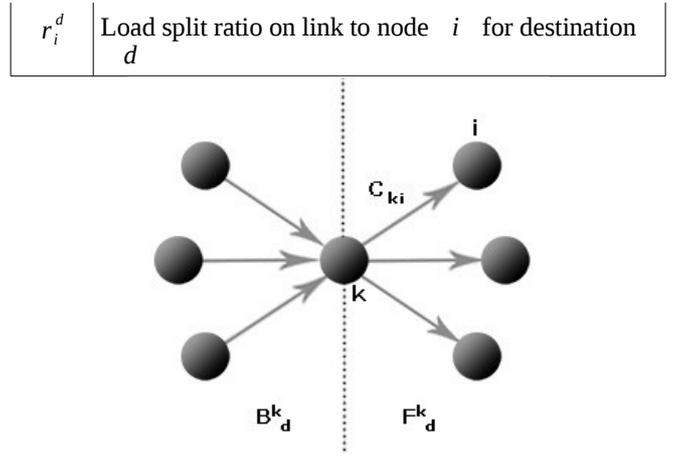

Fig 1: Forward and Backward sets of node $k$ for destination $d$

$F_d^i$ of $i$ (mutual movement).

- When a network is started, $F_d^k$ at all the nodes is empty. Consequently, at all the nodes, $B_d^k$ has all the neighboring nodes.
- Traffic from a node to a destination is always sent to the nodes in the forward set for that destination. Traffic is never sent in the backward direction making the algorithm *loop-free* at every instant.

### B. Computing Cost to Destination

Consider a node $k$ and a destination $d$ with $i \epsilon F_d^k$ as shown in Fig 1. For any link $(k,i)$ with link capacity $C_{ki}$, utilization is given by,

$$Utilization(u_{ki}) = \frac{Load\ on\ the\ link(k,i)}{Capacity\ of\ the\ link(k,i)} \quad (1)$$

Since the algorithm is made to work in dynamically changing scenarios, the link cost metric should be sensitive to congestion of links. The metric used here to model congestion is *available capacity* which for any link is given by,

$$Available\ capacity\ of\ link(k,i) = (1-u_{ki})C_{ki} \quad (2)$$

The total capacity from source $k$ to destination $d$ is then calculated using the available capacity of all the links along the path and is given as:

$$C_k^d = \sum_{i \in F_d^k} C_{ki}^d \quad (3)$$

$$C_{ki}^d = \begin{cases} min((1-u_{ki})C_{ki}, C_i^d) & ; \ i \neq d \\ (1-u_{ki})C_{ki} & ; \ i = d \end{cases} \quad (4)$$

For every destination $d$, node $k$ announces these capacities to its neighbor nodes depending upon whether they are in forward or backward set of $d$. Capacity announced in the backward links by the node $k$ is

$$_B C_k^d = C_k^d \quad (5)$$

In the forward links to node $i$, where $i$ is in $F_d^k$, the announced capacity is

$$_F C_{ki}^d = C_k^d - C_{ki}^d \quad (6)$$

This is poisoned reverse equivalent in multipath routing as the capacity through forward node $i$, $C_{ki}^d$, is not taken into account while announcing it the total capacity, $_FC_{ki}^d$, for destination $d$.

These announced capacities by $k$ are used by nodes in set $N_k$ to update their capacities to reach destination $d$. If the node is in set $B_d^k$, it updates its total capacity to $d$ via $k$ to $_BC_k^d$. If the node is in set $F_d^k$, it uses this capacity $_FC_{ki}^d$ to request node $k$ to move to its forward set for destination $d$. A node $l \in B_d^k$ is moved to forward link w.r.t. $k$ if

$$min(_FC_{lk}^d, C_{kl}(1-u_{kl})) > KC_k^d \qquad (7)$$

Here, $K$ is a control variable and can be taken as 10% (0.1). When node $l$ is moved to forward set of node $k$, $k$ is simultaneously moved to backward set of node $l$.

*C. Splitting Traffic*

All the traffic from node $k$ to destination $d$ is split statistically among all forward links with ratios $r_i^d$ on link to node $i$ where $r_i^d$ is given by,

$$r_i^d = \frac{C_{ki}^d}{C_k^d}; \qquad r_i^d \leq 1, \qquad \forall i \in F_d^k \qquad (8)$$

*D. Routing Table Structure*

Every node maintains two tables – neighbour table and main table. The *Neighbor Table* at any node $k$ stores the characteristics of all the neighboring links of node $k$. Each entry in this table is a triplet $[j, C_{kj}, u_{kj}]$ where $j \in N_k$.

The *Main Table* at any node $k$ stores, for each destination $d$, the forward set $F_d^k$ and the capacities and $C_k^d = \sum_{i \in F_d^k} C_{ki}^d$ for all $i \in F_d^k$.

*E. Packet Types*

The algorithm uses four types of packets. *Hello* packets are sent periodically by a node to identify and maintain neighbor relationships. Each *Hello* packet need to be acknowledged by sending another *Hello* packet in reverse direction. They can – a) contain the information about the link capacity $C_{kj}$, b) can be used to estimate the link capacity if it is not already programmed in the routers.

A coordination packet – *ForwardMoveRequest* is sent by any node $k$ to any node not in set $F_d^k$ but in $N_k$ (thus it is in $B_d^k$), requesting it to move to the set $F_d^k$. In this process it is ensured that every node except the destination must have at least one forward node for the destination. This is true for all nodes as destination. The movement is done when it will increase the capacity for the initiating node to destination thus improving the performance.

A coordination packet – *ForwardMoveResponse* packet is sent as a response to the *ForwardMoveRequest* packet. In this packet, a node may accept or reject the *ForwardMoveRequest* if this does not reduce the capacity to destination to unacceptable level.

*NeighborUpdate* packets are sent by every node to its neighbors and it contains the capacities from it to reach all the destinations; for a destination $d$, either $_BC_k^d$ or $_FC_{ki}^d$ will be announced to a neighbor. The received updates from neighbors are used along with the estimated utilization and link capacities to find the fraction to be used for splitting the traffic for a destination, through every link in the $F_d^k$ set. The *NeighborUpdate* packets are sent periodically or when a local table change happens due to the change in link status or due to *NeighborUpdate* received from any of the neighboring nodes.

*F. Timers*

There are five timers associated with this algorithm. *HelloTimer* at each node goes off every $T_H$ seconds (e.g. $T_H$ = 15 secs). Each time the timer expires, *Hello* packets are generated and sent to the neighbors and the timer is reset. The *Hello* packets need to be acknowledged by sending *Hello* packets in reverse. Every time a *Hello* packet is received, the *NeighborRemove* timer is reset. The neighbor is removed from the neighbor table, if the *NeighborRemove* timer expires. Typical value of *NeighborRemove* timer is 2 to 4 times the *HelloTimer* period.

*NeighborUpdateTimer* at each node goes off every $T_U$ seconds (e.g. $T_U$ = 30 secs). Each time the timer expires, *NeighborUpdate* packets are generated and sent to neighbors and *NeighborUpdateTimer* is reset. On receiving an update from a neighbor, *Timeout* timer is reset. If the updates from any neighboring node $j$ are not received, *NeighborRemove* timer is not expired and *Timeout* expires, then the capacity $C_{kj}^d$ is taken as zero for all $d$. It may be noted that a node keep on sending the neighbor updates to a node if *NeighborRemove* timer is not expired by *Timeout* expires. Typical value of *Timeout* is 5 – 6 times neighbor update periods.

Whenever a *ForwardMoveRequest* packet is sent, a *Move* timer is started at the sending node. The *ForwardMove-Response* packet must be received before this timer expires otherwise the *ForwardMoveRequest* packet is considered lost and a new *ForwardMoveRequest* packet is sent. This timer has a value of $T_M$ (e.g. $T_M$ = 30 secs).

III. THE ALGORITHM

- When a network comes up, routing tables at a node $k$ are initialized with the sets $N_k$ and $F_d^k$ as null. The $C_{kj}$ will be either estimated during the *Hello* exchange or can also be programmed in the router during the link installation time. The record for capacities $C_{ki}^d$ and $C_k^d$ will not be there and will be created as the *Hello* exchanges and routing table updates are received during network operations. This is done for all the nodes.

- The *Hello* packets are exchanged between each pair of neighbors $(k, j)$ utilization of each link $u_{kj}$ is calculated as ratio of time for which queue was not empty towards this neighbor to the observation time. These values of $u_{kj}$ and $C_{kj}$ along with $N_k$ fill up the entries of the neighbor tables.

- For every destination $d$, any node $k$ decide upon the forward set $F_d^k$ by the exchange of *ForwardMoveRequest* and *ForwardMoveResponse* packets. All the nodes connected to a destination $d$ directly, will maintain $d$ in their $F_d^k$ set, and $d$ will maintain all its neighbors in $B_d^d$ set. It shall be noted that $F_d^d$ set will always be empty.

- Each node $k$ will also calculate capacities $C_{ki}^d$ and $C_k^d$ which together with $F_d^k$ fill up the entries of main tables.

- Each node $k$ will announce $_FC_{ki}^d$ capacity to all the nodes $i$ in set $F_d^k$ and $_BC_k^d$ capacity to all the nodes in set $B_d^k$ by exchanging *NeighborUpdate* packets which will contain these announced capacities. The main table is recomputed at each node according to the received updates (the capacities $C_{ki}^d$ and $C_k^d$ are computed). If change happens in main table, node $k$ sends *NeighborUpdate* packets to all its neighbors in set $N_k$.

- After some time, each node will have at least one forward link for each destination. The number of forward links will be increased when moving a backward node to the forward set will increase the capacity for the initiating node to destination at least by fraction $K$.

- The amount of traffic for a destination $d$ routed on a link to node $i$ is decided by $r_i^d$ as given by (8).

IV. SIMULATION RESULTS

The simulations discussed in this section demonstrate the significant improvements achieved by our algorithm over single path and multipath algorithms. Instead of simulating any specific shortest path algorithm, we opted to restrict our algorithm to use only the best successor for packet forwarding. We use the label 'SP' for single-path routing in the graphs. For multipath algorithm, we used ECMP which again was implemented by restricting our algorithm to use equal cost paths to destination. The various parameters used for comparison are delay distribution, average delay, distribution of link utilization, average link utilization and throughput which are discussed in the following section in detail. All the comparisons are performed under the identical topological and traffic environments.

We performed simulations on the topology shown in Fig. 2. Network 1 has a connectivity that is high enough to ensure the existence of multiple paths and small enough to prevent a number of one-hop paths. We restricted the link capacities to a maximum of 10 Mbps, so that it becomes easy to sufficiently load the network. The load is assumed to be poisson in nature and service time of each node as exponentially distributed random variable. For simplicity, the topology is assumed to be stable (links or nodes do not fail) in all the situations. The plots of our algorithm are labeled with 'MP'. The experimental results are shown as symbols on these plots and are joined using lines.

*A. Delay Distribution*

Variable-sized packets are sent between (0 , 2) source-destination pair. The arrival rate of packets at node 0 is taken as 23 packets/sec and service rate of each node along the path as 25 packets/sec. Fig. 3 shows the distribution of end-to-end delay: delay from initial transmission of packet from source until packet is received at destination. From the plots we see that the distribution profile of MP is limited to lowest values of delays with that of ECMP and SP stretching over to higher values on x-axis. For the same number of packets delivered to the destination, we observe that MP delivers the largest number of packets with minimum delay of 0.2 secs as compared to SP and ECMP. This is due to the tendency of MP routing strategy which is to route packets on least loaded links and hence minimizing the queuing delay.

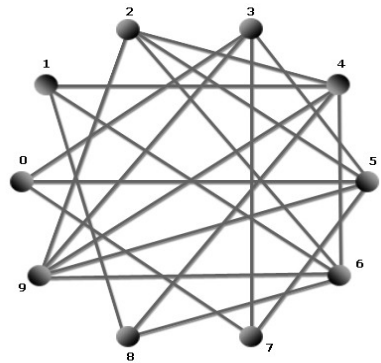

Fig 2: Network 1

*B. Average Delay*

Again the packets are sent between (0 , 2) and the arrival rate at node 0 is varied from 1 packet/sec to 24 packets/sec with the same departure rate of 25 packets/sec as above. We measured the effect of change in network load on the average delay of packets and the results are plotted in Fig. 4. We observe that under light load conditions the performance of all the algorithms is comparable but under high load environments delay of SP shoots upto 17 times of MP and that of ECMP to 4 times of MP. Specifically, the improvements in average packet delay ranges from 1.5 - 17 times over SP to about 1.3 - 4 times over ECMP under varying traffic conditions. Hence MP routing offers significant advantage over SP and ECMP.

*C. Throughput*

Fig. 5 compares the throughput of MP, SP and ECMP under the offered traffic conditions. As can be seen, under light load environments all the algorithms offer the same throughput but as load keeps on increasing, MP offers maximum utilization of network capacity as compared to SP and ECMP which attain their maximum possible throughput at much lower rates. From this graph the conclusion can be drawn that for the same performance, MP equires lesser total capacity of the network or for the same amount of total utilized capacity, MP performs better.

*D. Distribution of Link Utilization*

Poisson traffic is generated between every node to every other node in the network and distribution of link utilization values is plotted as given by Fig. 6. We observe that area under the curve of MP is maximum where all the links in the network are utilized to route packets. ECMP curve has an an area of 52% of MP meaning only 52% of total links are used and this value drops further to 47% in SP. Furthermore, in MP, utilization range varies from 0.1 - 0.5 with most of the

*E. Average Link Utilization*

Traffic is generated between every node pair in the network with arrival rates varying from 1 packet/sec to 14

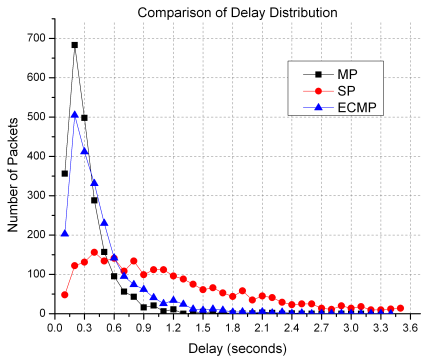

Fig 3: Delay Distribution

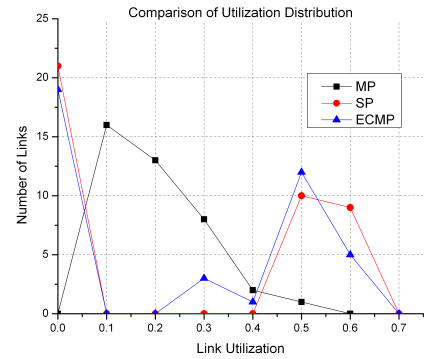

Fig 6: Distribution of Link Utilization

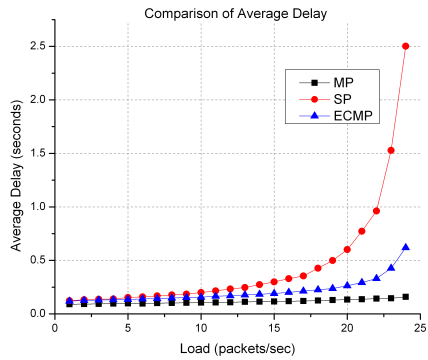

Fig 4: Average Packet Delay vs Load

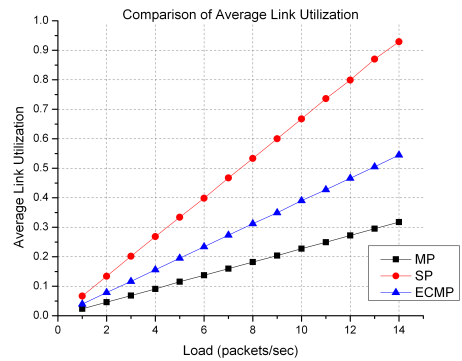

Fig 7: Average Link Utilization vs Load

packets/sec and a constant departure rate of 15 packets/sec. Fig. 7 provides the obtained results. Observe that average utilizations of SP are as much as 3 - 4 times those of MP routing and the same of ECMP as 1.5 - 2 times. These lowest utilization values indicates the reduction of congestion in MP which it achieves by accounting the bandwidth in choosing the multiple paths and splitting traffic among them.

## V. CONCLUSIONS

We have presented a new algorithm based on distance-vector routing which is free from count-to-infinity problem, provides multiple paths to destination that need not have equal costs and splits traffic among them in the ratios that need not be equal. The novelty of the algorithm lies in the formation of forward - backward sets at each node for every possible destination which ensures loop freedom and assignation of weights to forward nodes (next hops) which ensures optimal traffic split among them. We have shown through simulations that our proposed algorithm performs significantly better than SP and ECMP, in terms of per-packet delay, average delay, throughput and link utilization, under a wide range of different traffic conditions.

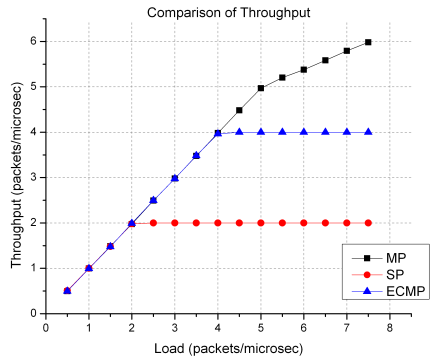

Fig 5: Throughput s Load

links utilized only 10% times and only a few of them having higher utilization values in range 0.4 - 0.5. In SP all the links are utilized in range 0.5 - 0.6, in ECMP the total no. of links in this range drops down with some of them having lesser utilization in range 0.3 - 0.4. The utilization distribution profile of MP is unimodal, of SP is bimodal and of ECMP is trimodal. The above mentioned values of utilization and their distribution pattern lead us to the conclusion that MP provides a more efficient flow distribution by balancing loads on all the links which results in their decreased utilization instead of loading only best paths in the network as is done by SP and ECMP.